# A Nonlinear Tracking Algorithm with Range-rate Measurements Based on Unbiased Measurement Conversion

Lianmeng Jiao, Quan Pan, Yan Liang, Feng Yang
School of Automation
Northwestern Polytechnical University
Xi'an, P. R. China
jiaolianmeng@mail.nwpu.edu.cn

*Abstract*—**The three-dimensional CMKF-U with only position measurements is extended to solve the nonlinear tracking problem with range-rate measurements in this paper. A pseudo measurement is constructed by the product of range and range-rate measurements to reduce the high nonlinearity of the range-rate measurements with respect to the target state; then the mean and covariance of the converted measurement errors are derived by the measurement conditioned method, showing better consistency than the transitional nested conditioning method; finally, the sequential filter was used to process the converted position and range-rate measurements sequentially to reduce the approximation error in the second-order EKF. Monte Carlo simulations show that the performance of the new tracking algorithm is better than the traditional one based on CMKF-D.**

## I. INTRODUCTION

In active sonar and radar systems the measurements of the position of a target is reported in polar or spherical coordinates (its range and azimuth or bearing (as well as elevation angle in 3D radar) with respect to the sensor location). However, the target motion is linearly modeled in Cartesian coordinates. In this case, tracking in Cartesian coordinates using sensor measurements is a nonlinear state estimation problem. To solve this problem, a basic idea is to convert the spherical measurements to Cartesian coordinates using the familiar nonlinear mapping between the two coordinate systems, yielding a pseudo-measurement model for classical Kalman filtering. This is called the converted measurements Kalman filtering (CMKF) method [1].

As shown in [2], the nonlinear transformation of unbiased spherical measurements to raw Cartesian converted measurements creates a bias in the converted measurements error. Debiasing the converted measurements is hence necessary for unbiased estimate based on Kalman filtering. Lerro and Bar-Shalom [3] firstly studied explicit solutions for the mean and covariance of the converted 2D measurements and presented a debiased CMKF (CMKF-D) algorithm which provides more accurate state estimate than EKF and traditional CMKF. Suchomski [4] extended the CMKF-D algorithm to the 3D spherical measurements. The unbiased CMKF (CMKF-U) subsequently developed by Mo and Bar-Shalom [5] obtained an unbiased converted measurement through multiplying the raw converted measurement by a vector of bias-elimination factors. However, in [3-5], the range-rate measurement is not considered though it is widely available [7]. In other words, the accuracy of tracking is possibly further improved by using range-rate measurements.

To solve the radar target tracking problem with range-rate measurements, in [7], the product of the range and range rate measurements is used as a pseudo measurement to reduce the nonlinearity of the range-rate measurement with respect to the target state, then statistics of the converted measurement errors of this pseudo measurement are obtained by the linearization method. But as exploited in [3], the converted measurement obtained by the linearization method is not consistent with its first two moments. Recently, the RCMKF-D algorithm [8] was proposed by Duan, Han, and Li to solve the radar target tracking problem with range-rate measurements through extending the CMKF-D algorithm with only position measurements in [4]. However, as indicated in [5] and [9], the CMKF-D algorithm which employs the nested conditioning method gives slightly biased estimates due to its nested conditional expectation operators. Meanwhile, recent research in [10] and [11] shows that the CMKF-U algorithm with only position measurements in [5] is measurement-conditioned through directly deriving the mean and covariance of the converted measurement errors conditioned on the sensor measurements, and as a result, has better consistency and robustness for noise distribution compared with CMKF-D.

To use the range-rate measurement more sufficiently, the RCMKF-U algorithm is proposed in this paper which extends the 3D CMKF-U with only position measurements in [5] to solve the nonlinear tracking problem with range-rate measurements. A pseudo measurement is constructed by the product of range and range-rate measurement to reduce the high nonlinearity of the range-rate measurement with respect to the target state; then the mean and covariance of the converted measurement errors are derived by the measurement conditioned method, showing better consistency than the nested conditioning method; finally, since the pseudo measurement is

This work is supported by the National Natural Science Foundation of China (61135001, 610741179), the Basic Research Foundation of NWPU (JC201015), and the Soaring Star Plan of NWPU.

quadratic in the target state, the Cholesky factorization is used to decorrelate the converted position and pseudo measurement errors, thus the position and pseudo measurements can be sequentially processed to reduce the approximation error in the second-order EKF further.

The rest of this paper is organized as following. Section II presents the target dynamic model and radar measurement equation. Section III derives the statistics of the converted measurement errors conditioned on sensor's spherical measurements and gives the consistency analysis. Section IV describes the sequential filtering with the converted measurements. Numerical simulations are performed in Section V and conclusions are presented in Section VI.

## II. PROBLEM FORMULATION

### A. Target Dynamic Model

The target dynamic model can be modeled in Cartesian coordinates as

$$X_{k+1} = \Phi_k X_k + G_k U_k + \Gamma_k W_k, \tag{1}$$

where, $X_k = [x_k \; y_k \; z_k \; \dot{x}_k \; \dot{y}_k \; \dot{z}_k \; s_{1\times(n-6)}]^T \in \mathbb{R}^n$ is the state vector consisting of the position and corresponding velocity components along the $x$, $y$ and $z$ coordinates at time step $k$, respectively. $s_{1\times(n-6)}$ is other state components such as acceration. $\Phi_k \in \mathbb{R}^{n\times n}$ is the state transition matrix; $U_k$ is the deterministic input matrix; $W_k$ is the zero-mean white Gaussian process noise with known covariance $Q_k$. $G_k$ and $\Gamma_k$ are the known matrices with appropriate dimensions.

### B. Radar Measurement Equation

In the spherical coordinates, we assume that the Doppler radar locates at the origin of Cartesian coordinates. The radar measurement equation with range-rate measurements can be expressed as

$$\begin{aligned} Z_k^m &= \left[r_k^m, \theta_k^m, \varphi_k^m, \dot{r}_k^m\right]^T \\ &= f(X_k) + V_k^m \\ &= \left[r_k, \theta_k, \varphi_k, \dot{r}_k\right]^T + \left[\tilde{r}_k, \tilde{\theta}_k, \tilde{\varphi}_k, \tilde{\dot{r}}_k\right]^T, \end{aligned} \tag{2}$$

where

$$\begin{cases} r_k = \sqrt{x_k^2 + y_k^2 + z_k^2} \\ \theta_k = \arctan(y_k / x_k) \\ \varphi_k = \arctan\left(z_k \Big/ \sqrt{x_k^2 + y_k^2}\right) \\ \dot{r}_k = (x_k\dot{x}_k + y_k\dot{y}_k + z_k\dot{z}_k) \Big/ \sqrt{x_k^2 + y_k^2 + z_k^2} \end{cases}.$$

$Z_k^m$ is the measurement vector obtained from radar at time step $k$, which consists of the rang $r_k^m$, the bearing $\theta_k^m$, the elevation $\varphi_k^m$, and the range rate $\dot{r}_k^m$. $V_k^m$ is the corresponding measurement noise vector, the elements of which are all assumed to be zero-mean white Gaussian noises with known variances $\sigma_r^2$, $\sigma_\theta^2$, $\sigma_\varphi^2$ and $\sigma_{\dot{r}}^2$, respectively. It is assumed that $\tilde{r}_k$, $\tilde{\theta}_k$, and $\tilde{\varphi}_k$ are statistically independent, and $\tilde{r}_k$ and $\tilde{\dot{r}}_k$ are correlated with correlation coefficient $\rho$.

## III. MEASUREMENT CONVERSION

### A. Measurement Conversion with Range-rate Measurements

The position measurements (range, bearing and elevation) in spherical coordinates can be transformed into the pseudo-linear form in the Cartesian coordinates by

$$\begin{cases} x_k^c = r_k^m \cos\varphi_k^m \cos\theta_k^m = x_k + \tilde{x}_k \\ y_k^c = r_k^m \cos\varphi_k^m \sin\theta_k^m = y_k + \tilde{y}_k \\ z_k^c = r_k^m \sin\varphi_k^m = z_k + \tilde{z}_k \end{cases}, \tag{3}$$

where $\tilde{x}_k$, $\tilde{y}_k$, $\tilde{z}_k$ are the position conversion measurement errors along $x$, $y$, $z$ directions in Cartesian coordinates, respectively.

To reduce the strong nonlinearity between the range-rate measurement and the target state, as in [7], the following pseudo measurement conversion equation can be utilized

$$\eta_k^c = r_k^m \dot{r}_k^m = x_k\dot{x}_k + y_k\dot{y}_k + z_k\dot{z}_k + \tilde{\eta}_k, \tag{4}$$

where $\tilde{\eta}_k$ is the converted pseudo measurement error in the Cartesian coordinates.

From (3) to (4), conversion of the radar measurements with rang rate from the spherical coordinates to the Cartesian coordinates can be totally expressed as

$$\begin{aligned} Z_k^c &= [x_k^c, y_k^c, z_k^c, \eta_k^c]^T = h_k(X_k) + V_k^c \\ &= [x_k, y_k, z_k, x_k\dot{x}_k + y_k\dot{y}_k + z_k\dot{z}_k]^T + [\tilde{x}_k, \tilde{y}_k, \tilde{z}_k, \tilde{\eta}_k]^T. \end{aligned} \tag{5}$$

### B. Statistics of Converted Measurement Errors Conditioned on the Spherical Measurements

In RCMKF-D algorithm [8], the nested conditioning method is used to get the statistics of converted measurement errors, which is to find the mean and covariance conditioned on the unknown ideal measurement $Z_k = \left[r_k, \theta_k, \varphi_k, \dot{r}_k\right]^T$ first (i.e., $\mu_k^t = E[V_k^c \mid Z_k]$ and $R_k^t = \mathrm{cov}[V_k^c \mid Z_k]$ ) and then find their expectations conditioned on the noisy measurement $Z_k^m = [r_k^m, \theta_k^m, \varphi_k^m, \dot{r}_k^m]^T$ (i.e., $\mu_k^a = E[\mu_k^t \mid Z_k]$ and $R_k^a = E[R_k^t \mid Z_k]$ ). However, this method gives slightly biased estimates due to its nested conditional expectation operators and has a consistency problem for large noise distribution.

This part derives explicit expressions for the converted measurement errors mean and covariance, when those quantities are conditioned on the spherical measurements directly.

Using (2) and (5), we can derive the practical bias and covariance of the converted measurement errors conditioned on the spherical measurements with range rate as

$$\mu_k^c = E\left[V_k^c \mid r_k^m, \theta_k^m, \varphi_k^m, \dot{r}_k^m\right] = [\mu_k^x, \mu_k^y, \mu_k^z, \mu_k^\eta]^T \tag{6}$$

$$R_k^c = \operatorname{cov}\left[V_k^c \mid r_k^m, \theta_k^m, \varphi_k^m, \dot{r}_k^m\right] = \begin{bmatrix} R_k^{xx} & R_k^{xy} & R_k^{xz} & R_k^{x\eta} \\ R_k^{xy} & R_k^{yy} & R_k^{yz} & R_k^{y\eta} \\ R_k^{xz} & R_k^{yz} & R_k^{zz} & R_k^{z\eta} \\ R_k^{x\eta} & R_k^{y\eta} & R_k^{z\eta} & R_k^{\eta\eta} \end{bmatrix} \tag{7}$$

where $\mu_k^x$, $\mu_k^y$, $\mu_k^z$ and $R_k^{xx}$, $R_k^{yy}$, $R_k^{zz}$, $R_k^{xy}$, $R_k^{xz}$, $R_k^{yz}$ are expressed as (8) – (14). (The same as in [10])

$$\begin{cases} \mu_k^x = r_k^m \cos\theta_k^m \cos\varphi_k^m (1 - \lambda_\theta \lambda_\varphi) \\ \mu_k^y = r_k^m \sin\theta_k^m \cos\varphi_k^m (1 - \lambda_\theta \lambda_\varphi) \\ \mu_k^z = r_k^m \sin\varphi_k^m (1 - \lambda_\varphi) \end{cases} \tag{8}$$

$$R_k^{xx} = -\lambda_\theta^2 \lambda_\varphi^2 (r_k^m)^2 \cos^2\theta_k^m \cos^2\varphi_k^m + \frac{1}{4}((r_k^m)^2 + \sigma_r^2)(1 + \lambda_\theta^{'} \cos 2\theta_k^m)(1 + \lambda_\varphi^{'} \cos 2\varphi_k^m) \tag{9}$$

$$R_k^{yy} = -\lambda_\theta^2 \lambda_\varphi^2 (r_k^m)^2 \sin^2\theta_k^m \cos^2\varphi_m + \frac{1}{4}((r_k^m)^2 + \sigma_r^2)(1 + \lambda_\theta^{'} \cos 2\theta_k^m)(1 + \lambda_\varphi^{'} \cos 2\varphi_k^m) \tag{10}$$

$$R_k^{zz} = -\lambda_\varphi^2 (r_k^m)^2 \sin^2\varphi_k^m + \frac{1}{2}((r_k^m)^2 + \sigma_r^2)(1 - \lambda_\varphi^{'} \cos 2\varphi_k^m) \tag{11}$$

$$R_k^{xy} = -\lambda_\theta^2 \lambda_\varphi^2 (r_k^m)^2 \sin\theta_k^m \cos\theta_k^m \cos^2\varphi_k^m + \frac{1}{4}((r_k^m)^2 + \sigma_r^2)\lambda_\theta^{'} \sin 2\theta_k^m (1 + \lambda_\varphi^{'} \cos 2\varphi_k^m) \tag{12}$$

$$R_k^{xz} = -\lambda_\theta \lambda_\varphi^2 (r_k^m)^2 \cos\theta_k^m \sin\varphi_k^m \cos\varphi_k^m + \frac{1}{2}((r_k^m)^2 + \sigma_r^2)\lambda_\theta \lambda_\varphi^{'} \cos\theta_k^m \sin 2\varphi_k^m \tag{13}$$

$$R_k^{yz} = -\lambda_\theta \lambda_\varphi^2 (r_k^m)^2 \sin\theta_k^m \sin\varphi_k^m \cos\varphi_k^m + \frac{1}{2}((r_k^m)^2 + \sigma_r^2)\lambda_\theta \lambda_\varphi^{'} \sin\theta_k^m \sin 2\varphi_k^m \tag{14}$$

where,

$$\begin{cases} \lambda_\theta = E\left[\cos\tilde{\theta}_m\right] = e^{-\sigma_\theta^2/2} \\ \lambda_\theta^{'} = E\left[\cos 2\tilde{\theta}_m\right] = e^{-2\sigma_\theta^2} \\ \lambda_\varphi = E\left[\cos\tilde{\varphi}_m\right] = e^{-\sigma_\varphi^2/2} \\ \lambda_\varphi^{'} = E\left[\cos 2\tilde{\varphi}_m\right] = e^{-2\sigma_\varphi^2} \end{cases}.$$

The mean and covariance of converted pseudo measurement error $\mu_k^\eta$, $R_k^{x\eta}$, $R_k^{y\eta}$, $R_k^{z\eta}$, and $R_k^{\eta\eta}$ can be expressed as

$$\begin{cases} \mu_k^\eta = -\rho\sigma_r\sigma_{\dot{r}} \\ R_k^{x\eta} = \lambda_\theta \lambda_\varphi (\sigma_r^2 \dot{r}_k^m + r_k^m \rho\sigma_r\sigma_{\dot{r}}) \cos\varphi_k^m \cos\theta_k^m \\ R_k^{y\eta} = \lambda_\theta \lambda_\varphi (\sigma_r^2 \dot{r}_k^m + r_k^m \rho\sigma_r\sigma_{\dot{r}}) \cos\varphi_k^m \sin\theta_k^m \\ R_k^{z\eta} = \lambda_\varphi (\sigma_r^2 \dot{r}_k^m + r_k^m \rho\sigma_r\sigma_{\dot{r}}) \sin\varphi_k^m \\ R_k^{\eta\eta} = (r_k^m)^2 \sigma_{\dot{r}}^2 + \sigma_r^2 (\dot{r}_k^m)^2 + (1 + \rho^2)\sigma_r^2\sigma_{\dot{r}}^2 + 2 r_k^m \dot{r}_k^m \rho\sigma_r\sigma_{\dot{r}} \end{cases} \tag{15}$$

### *C. Consistency Analysis of New and the Nested Conditioning Method*

In this part the statistical consistency of the unbiased measurement conversion given above is tested compared with the nested conditioning method. In order to perform the consistency test, the following sample average of the normalized error squared (NES) [2, 6] associated with converted measurement errors is considered.

$$\bar{\psi} = \frac{1}{N} \sum_{i=1}^{N} \tilde{z}_i^T P_{\tilde{z}\tilde{z}}^{-1} \tilde{z}_i \tag{16}$$

where, $\tilde{z}_i$ denotes the vector of converted measurement errors in realization $i$, compensated for the hypothesized bias, $P_{\tilde{z}\tilde{z}}$ is the hypothesized covariance of the errors, and $N$ denotes the assumed number of test samples.

It should be noted that this statistical consistency check is based on a theorem in statistics that the sum of independent zero-mean unity-variance Gaussian variables squared has a chi-square distribution. So the evaluation here is approximate since the converted measurement errors are not independent and not Gaussian as well in general. Considering the measurements obtained from a 2D radar, which consists of the range $r_k^m$, the bearing $\theta_k^m$, and the range rate $\dot{r}_k^m$. The mean value of the statistic (16) is equal to 3 when there is no bias and the assumed hypothesized covariance is matched to the actual error covariance. If the errors are jointly Gaussian then the distribution of $N\bar{\psi}$ is chi-square with $3N$ degrees of freedom. Using $N = 1000$ samples of converted measurements one obtains the following acceptance region for the 99.8% probability bounds [6]

$$[\chi_{3000}^2(0.001), \chi_{3000}^2(0.999)] = [2.76 \times 1000, 3.24 \times 1000].$$

The assumed true object position is at $r = 10\text{km}$ with bearing $\theta = 45^\circ$ and the rang-rate is $\dot{r} = 100\text{m/s}$. The standard deviations of measurement errors are as follows: $\sigma_r = 100\text{m}$, $\sigma_{\dot{r}} = 5\text{m/s}$ and $\sigma_\theta \in [0^\circ, 30^\circ]$. The results of the consistency check in this case for the measurement conditioned conversion and the nested conditioning conversion are shown in Fig.l. It is easily see that the proposed conversion, is consistent (at least approximately since the errors are not Gaussian) even for relatively large measurement errors, while the nested conditioning conversion results in consistent characteristics only for a narrower range of measurement errors.

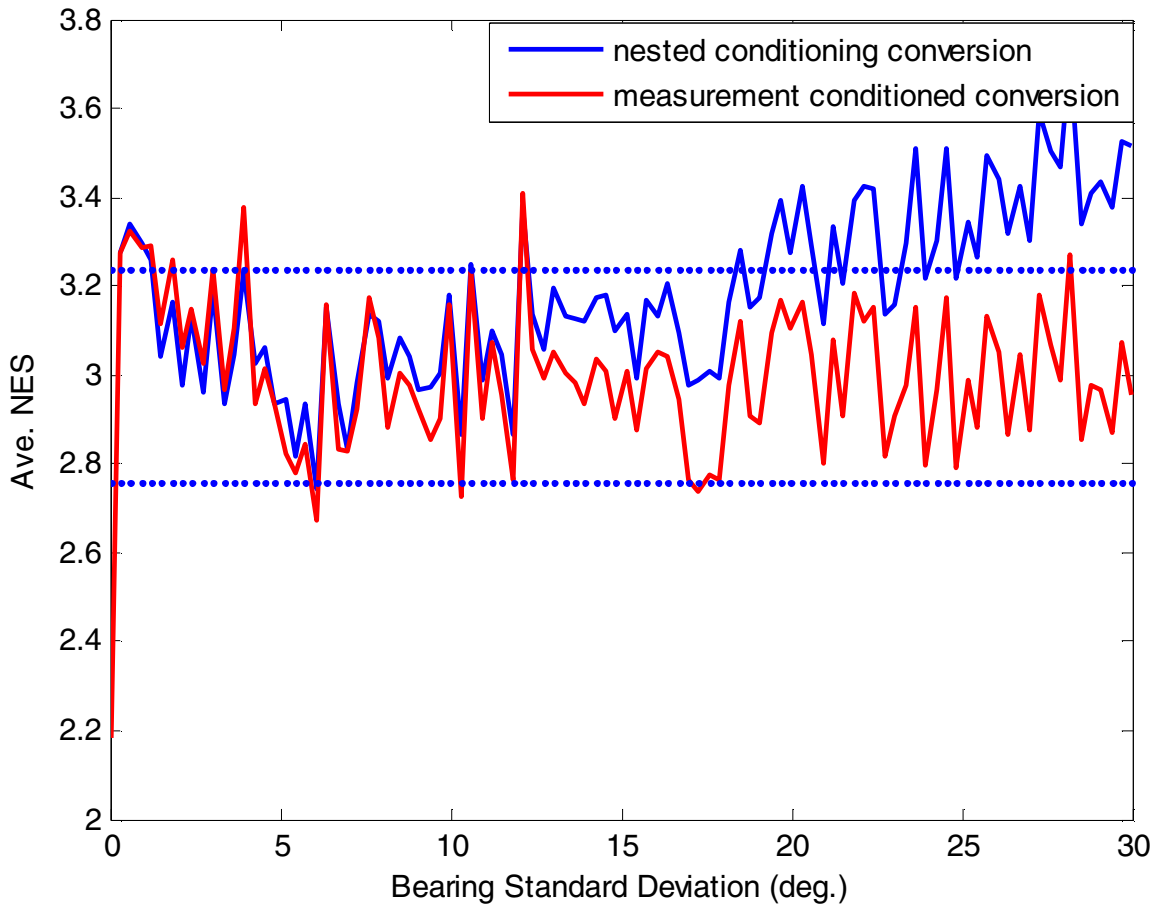

Figure 1. Average NES for measurement conditioned conversion ( $\mu_k^c$ , $R_k^c$ ) and nested conditioning conversion ( $\mu_k^a$ , $R_k^a$ ) evaluated at measured position.

## IV. TRACKING WITH CONVERTED MEASUREMENTS

It can be seen from (5) that the range-rate converted measurements are still nonlinear function of the target state, while the position converted measurements are linear functions of the target state. The sequential filter (SF) [12] was used to process position and range-rate measurements sequentially, that is, the position conversion measurements are processed first to obtain the target state estimate $\hat{X}_{k/k}^p$ , then the nonlinear function $h_k(X_k)$ is linearized by the Taylor series expansion around $\hat{X}_{k/k}^p$ , and thus the linearization errors should be reduced.

### *A. Decorrelation between Position and Pseudo Measurements*

From (7), the converted measurement errors of position and pseudo measurement $\eta_k^c$ are correlated, so they should be decorrelated first before sequential filtering.

Covariance matrix $R_k^c$ of the converted measurement errors can be rewritten as

$$R_k^c = \begin{bmatrix} R_k^{pp} & (R_k^{\eta p})^T \\ R_k^{\eta p} & R_k^{\eta\eta} \end{bmatrix}. \tag{17}$$

Set $L_k = -R_k^{\eta p}(R_k^{pp})^{-1} = \begin{bmatrix} L_k^1 & L_k^2 & L_k^3 \end{bmatrix}$, $B_k = \begin{bmatrix} I_3 & 0 \\ L_k & 1 \end{bmatrix}$.

Left-multiplying $B_k$ on both sides of (5), from Cholesky factorization [1], one can get

$$\begin{cases} Z_k^{c,p} = H_k^{c,p} X_k + V_k^{c,p} \\ \varepsilon_k^c = h_k^\varepsilon (X_k) + \tilde{\varepsilon}_k \end{cases}, \tag{18}$$

where,

$$\begin{cases} Z_k^{c,p} = \begin{bmatrix} x_k^c & y_k^c & z_k^c \end{bmatrix}^T \\ H_k^{c,p} = \begin{bmatrix} I_3 & 0_{3\times(n-3)} \end{bmatrix} \\ V_k^{c,p} = \begin{bmatrix} \tilde{x}_k & \tilde{y}_k & \tilde{z}_k \end{bmatrix}^T \\ E\left[V_k^{c,p}\right] = \mu_k^p = \begin{bmatrix} \mu_k^x & \mu_k^y & \mu_k^z \end{bmatrix}^T \\ \operatorname{cov}\left[V_k^{c,p}\right] = R_k^{pp} \\ \varepsilon_k^c = L_k^1 x_k^c + L_k^2 y_k^c + L_k^3 z_k^c + \eta_k^c \\ h_k^c(X_k) = L_k^1 x_k + L_k^2 y_k + L_k^3 z_k + x_k \dot{x}_k + y_k \dot{y}_k + z_k \dot{z}_k \\ \tilde{\varepsilon}_k = L_k^1 \tilde{x}_k + L_k^2 \tilde{y}_k + L_k^3 \tilde{z}_k + \tilde{\eta}_k \\ E\left[\tilde{\varepsilon}_k\right] = \mu_k^\varepsilon = L_k^1 \mu_k^x + L_k^2 \mu_k^y + L_k^3 \mu_k^z + \mu_k^\eta \\ \operatorname{cov}\left[\tilde{\varepsilon}_k\right] = R_k^{\varepsilon\varepsilon} = R_k^{\eta\eta} - R_k^{\eta p}(R_k^{pp})^{-1}(R_k^{\eta p})^T \end{cases} \tag{19}$$

### *B. SF with EKF*

The SF means that the Kalman filter (KF) is suited for the converted position measurements since its linear characteristic, the EKF, is used to update the mean and covariance from the KF by using the pseudo measurement.

#### *1) Position Measurement Filtering – KF Stage*

Time update and measurement update of the target state by position measurements $Z_k^{c,p}$ , can be implemented as follows.

$$\begin{cases} \hat{X}_{k/k-1} = \Phi_{k-1} \hat{X}_{k-1/k-1} + G_{k-1} U_{k-1} \\ P_{k/k-1} = \Phi_{k-1} P_{k-1/k-1} \Phi_{k-1}^T + \Gamma_{k-1} Q_{k-1} \Gamma_{k-1}^T \end{cases} \tag{20}$$

$$\begin{cases} K_k^p = P_{k/k-1}(H_k^{c,p})^T \left[ H_k^{c,p} P_{k/k-1} (H_k^{c,p})^T + R_k^{pp} \right]^{-1} \\ \hat{X}_{k/k}^p = \hat{X}_{k/k-1} + K_k^p \left[ Z_k^{c,p} - \mu_k^p - H_k^{c,p} \hat{X}_{k/k-1} \right] \\ P_{k/k}^p = (I_n - K_k^p H_k^{c,p}) P_{k/k-1} \end{cases} \tag{21}$$

#### *2) Update with Pseudo Measurement – EKF Stage*

From (18), the pseudo measurement $\varepsilon_k^c$ is a quadratic function of the target state, and so the nonlinear filtering estimation for the target state can be achieved by the second-order EKF in [7] as

$$\begin{cases} K_k^\varepsilon = P_{k/k}^p (H_k^\varepsilon)^T \left[ H_k^\varepsilon P_{k/k}^p (H_k^\varepsilon)^T + R_{k,a}^{\varepsilon\varepsilon} + A_k \right]^{-1} \\ \hat{X}_{k/k} = \hat{X}_{k/k}^p + K_k^\varepsilon \left[ \varepsilon_k^c - \mu_k^\varepsilon - h_k^\varepsilon(\hat{X}_{k/k}^p) - \frac{1}{2}\delta_k^2 \right] \\ P_{k/k} = (I_n - K_k^\varepsilon H_k^\varepsilon) P_{k/k}^p \end{cases}, \tag{22}$$

where, $H_k^\varepsilon$ is the Jacobian of $h_k^\varepsilon(X_k)$ around $\hat{X}_{k/k}^p$ , and

$$H_k^c = \begin{bmatrix} L_k^1 + \hat{x}_{k/k}^p & L_k^2 + \hat{y}_{k/k}^p & L_k^3 + \hat{z}_{k/k}^p \\ \hat{x}_{k/k}^p & \hat{y}_{k/k}^p & \hat{z}_{k/k}^p & 0_{1\times(n-6)} \end{bmatrix},$$

$\delta_k^2$ consists of the second-order derivative of $h_k^{\varepsilon}(X_k)$ and

$$\begin{aligned}
\delta_k^2 &= 2P_{k/k}^p(1,4) + 2P_{k/k}^p(2,5) + 2P_{k/k}^p(3,6), \\
A_k &= P_{k/k}^p(1,1)P_{k/k}^p(4,4) + P_{k/k}^p(2,2)P_{k/k}^p(5,5) + \\
&\quad P_{k/k}^p(3,3)P_{k/k}^p(6,6) + 2P_{k/k}^p(1,2)P_{k/k}^p(4,5) + \\
&\quad 2P_{k/k}^p(1,5)P_{k/k}^p(2,4) + 2P_{k/k}^p(1,3)P_{k/k}^p(4,6) + \\
&\quad 2P_{k/k}^p(1,6)P_{k/k}^p(3,4) + 2P_{k/k}^p(2,3)P_{k/k}^p(5,6) + \\
&\quad 2P_{k/k}^p(2,6)P_{k/k}^p(3,5) + [P_{k/k}^p(1,4)]^2 + \\
&\quad [P_{k/k}^p(2,5)]^2 + [P_{k/k}^p(3,6)]^2
\end{aligned}.$$

$P_{k/k}^p(i,j)$ represents the element located at the $i-$th row and $j-$th column of $P_{k/k}^p$.

## V. Simulation Results

In order to evaluate the performance of the new nonlinear tracking algorithm with range-rate measurements based on unbiased measurement conversion, we consider two typical test cases with different maneuvering characteristics with 500 Monte-Carlo runs. The target dynamic model in (1) is

$$X_{k+1} = \begin{bmatrix} 1 & 0 & T & 0 \\ 0 & 1 & 0 & T \\ 0 & 0 & 1 & 0 \\ 0 & 0 & 0 & 1 \end{bmatrix} X_k + \begin{bmatrix} T^2/2 & 0 \\ 0 & T^2/2 \\ T & 0 \\ 0 & T \end{bmatrix} \begin{bmatrix} w_k^x \\ w_k^y \end{bmatrix},$$

where, $X_k = [x_k, y_k, \dot{x}_k, \dot{y}_k]^T$, and the sampling interval $T = 1\text{s}$. $w_k^x$ and $w_k^y$ are zero-mean uncorrelated Gaussian white process noises with known standard deviation $0.01\text{m/s}^2$.

The sensor is assumed fixed at the origin of the polar coordinate and takes 100 range, bearing and range rate measurements. The corresponding measurement noise standard deviations are $\sigma_r = 200\text{m}$, $\sigma_\theta = 2.5^\circ$ and $\sigma_{\dot{r}} = 1\text{m/s}$, respectively. The correlation coefficient between the range and range-rate measurement noises is $\rho = 0.3$.

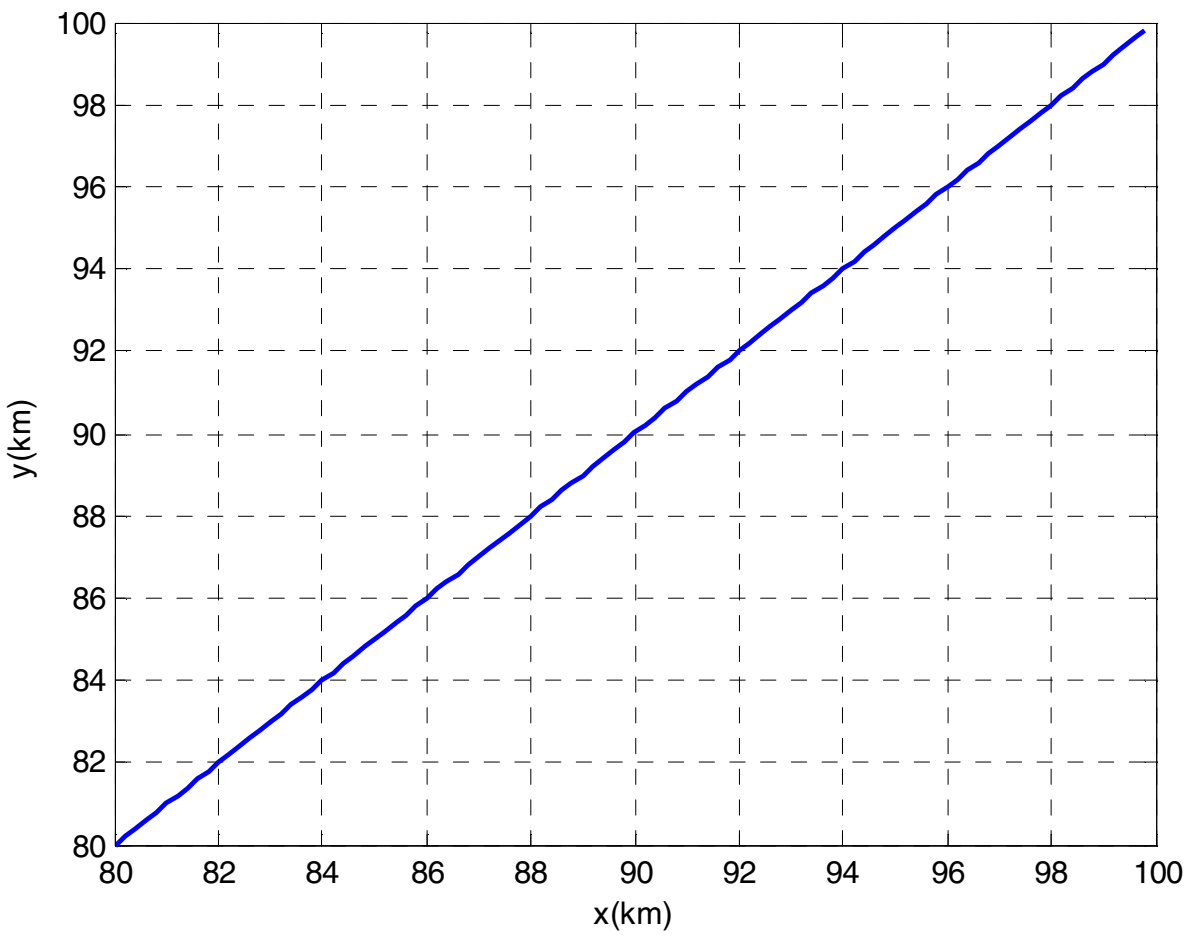

Figure 2. Target's real trajectory (Case 1)

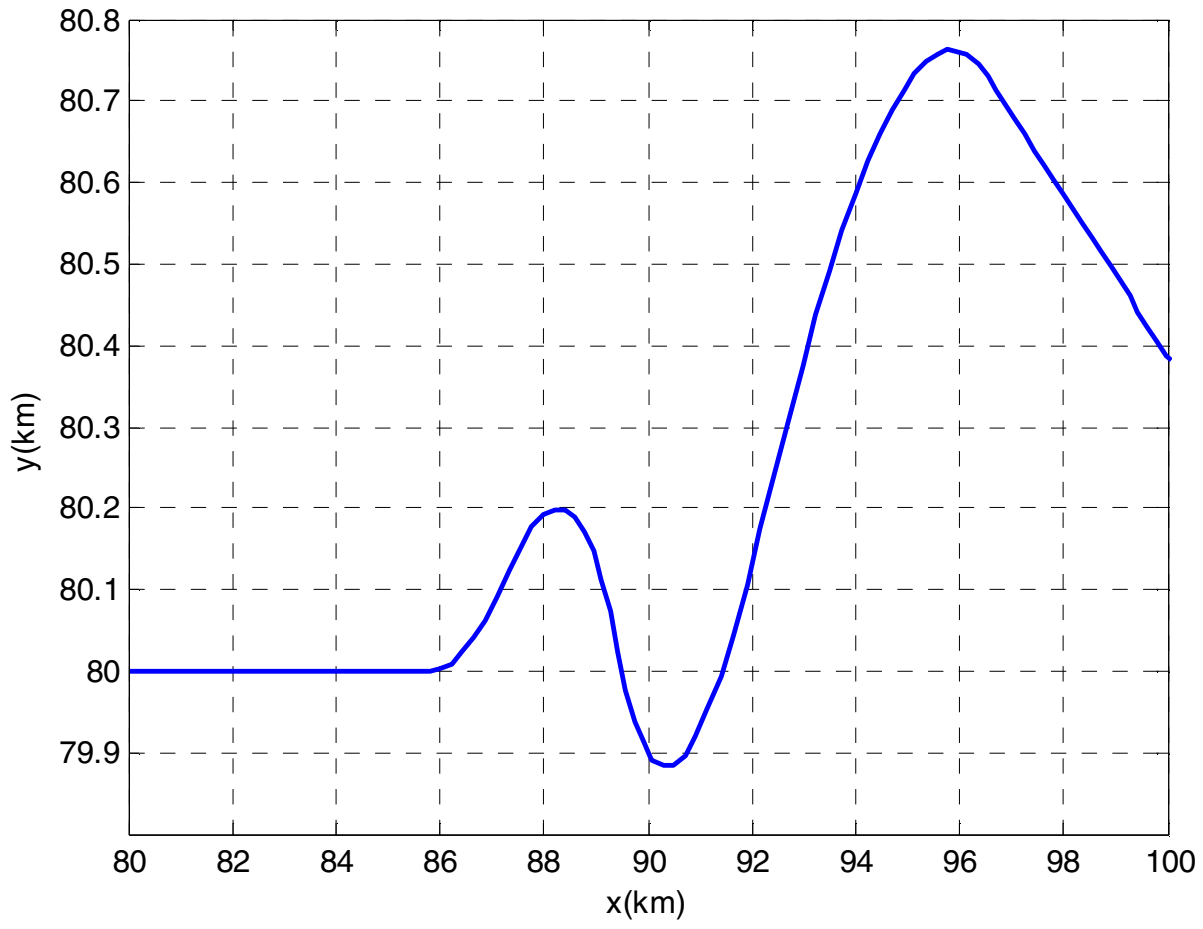

Figure 3. Target's real trajectory (Case 2)

*Case 1*: The target moves with nearly constant velocity. The target's initial location is (80km, 80km) and the initial velocity is (200m/s, 200m/s). The target's real trajectory is shown in Fig.2.

*Case 2*: The target moves with high maneuvers. The target's initial location is (80km, 80km) and the initial velocity is (0m/s, 200m/s). It maneuvers at 31s, 38s, 49s, 61s, 65s, 66s, and 81s with the acceleration of $5\text{m/s}^2$, $-8\text{m/s}^2$, $10\text{m/s}^2$, $0\text{m/s}^2$, $-10\text{m/s}^2$, $-5\text{m/s}^2$, and $0\text{m/s}^2$ respectively in both coordinate axes. The target's real trajectory is shown in Fig.3.

The position root-mean-squared errors (RMSE) of the proposed nonlinear tracking algorithm base on unbiased converted measurements (RCMKF-U), compared with the method based on debiased converted measurements (RCMKF-D) in [8], are shown in Fig.4-5.

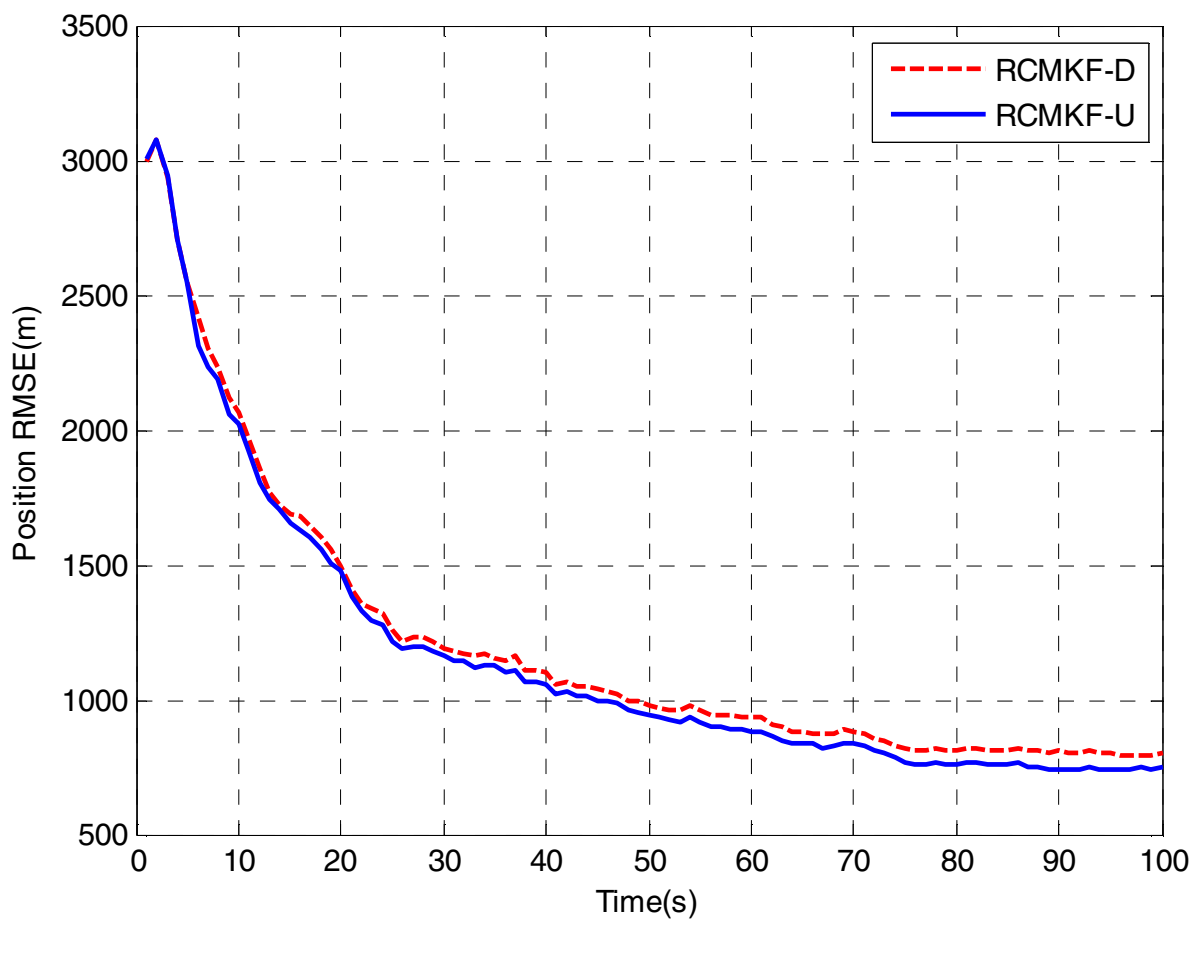

Figure 4. Target's Position RMSE (Case 1)

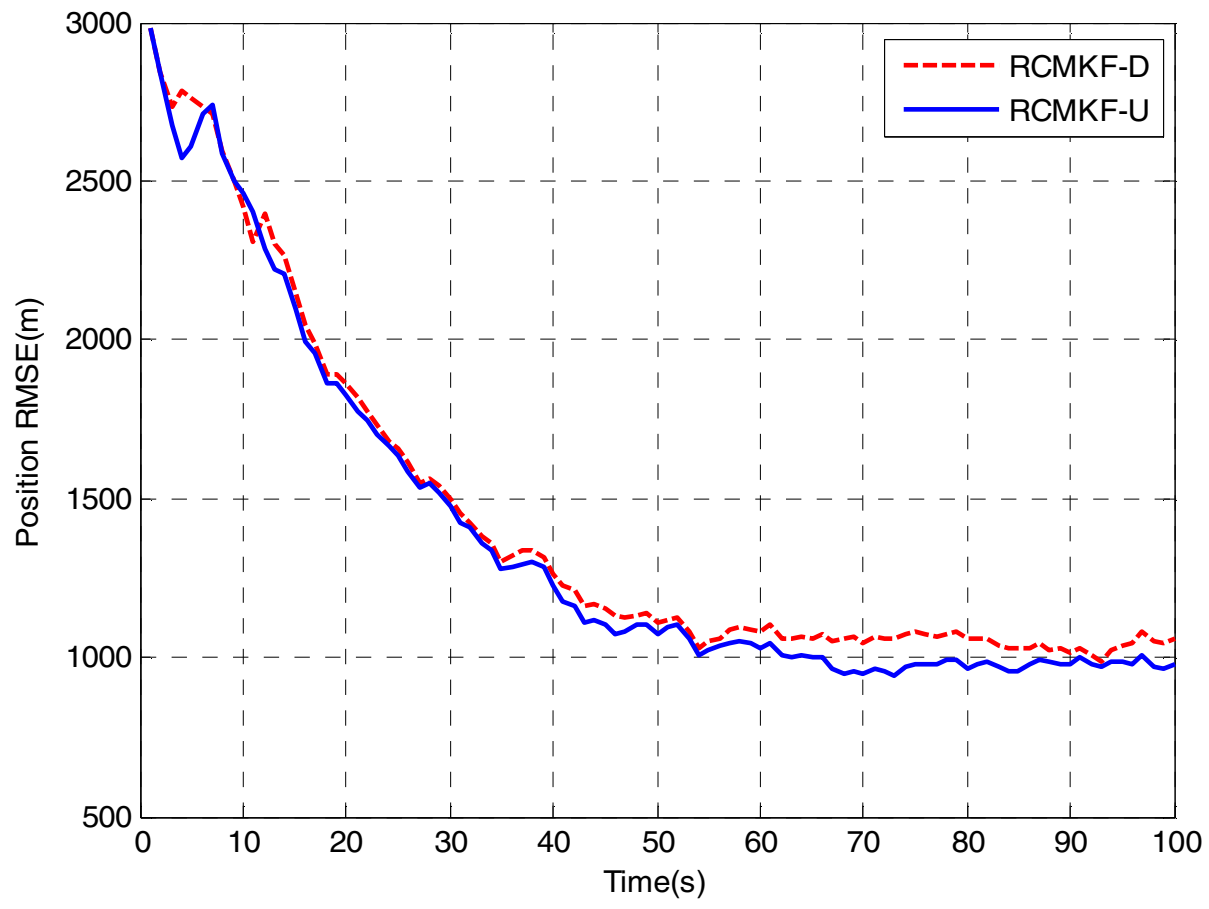

Figure 5. Target's Position RMSE (Case 2)

From the simulation results we can see that the proposed RCMKF-U is robust and performs better for both cases compared with RCMKF-D. This proves that the unbiased converted measurement improves the accuracy of nonlinear tracking filter.

## VI. CONCLUSIONS

A new nonlinear tracking algorithm with range-rate measurements is presented in this paper. Different from the traditional method, the consistent mean and covariance of the converted measurement errors are derived conditioned on the spherical measurements. Monte Carlo simulations show that the performance of the new tracking algorithm is better than the traditional one based on CMKF-D.